# Compressed Matrix Multiplication[*]

Rasmus Pagh
IT University of Copenhagen
pagh@itu.dk

## ABSTRACT
Motivated by the problems of computing sample covariance matrices, and of transforming a collection of vectors to a basis where they are sparse, we present a simple algorithm that computes an approximation of the product of two $n$-by-$n$ real matrices $A$ and $B$. Let $||AB||_F$ denote the Frobenius norm of $AB$, and $b$ be a parameter determining the time/accuracy trade-off. Given 2-wise independent hash functions $h_1, h_2 : [n] \rightarrow [b]$, and $s_1, s_2 : [n] \rightarrow \{-1, +1\}$ the algorithm works by first "compressing" the matrix product into the polynomial

$$p(x) = \sum_{k=1}^{n} \left( \sum_{i=1}^{n} A_{ik} s_1(i) \, x^{h_1(i)} \right) \left( \sum_{j=1}^{n} B_{kj} s_2(j) \, x^{h_2(j)} \right) \;.$$

Using FFT for polynomial multiplication, we can compute $c_0, \dots, c_{b-1}$ such that $\sum_i c_i x^i = (p(x) \bmod x^b) + (p(x) \text{ div } x^b)$ in time $\tilde{\mathcal{O}}(n^2 + nb)$. An unbiased estimator of $(AB)_{ij}$ with variance at most $||AB||_F^2 / b$ can then be computed as:

$$C_{ij} = s_1(i) \, s_2(j) \, c_{(h_1(i) + h_2(j)) \bmod b} \;.$$

Our approach also leads to an algorithm for computing $AB$ exactly, whp., in time $\tilde{\mathcal{O}}(N + nb)$ in the case where $A$ and $B$ have at most $N$ nonzero entries, and $AB$ has at most $b$ nonzero entries. Also, we use error-correcting codes in a novel way to recover significant entries of $AB$ in near-linear time.

## Categories and Subject Descriptors
F.2 [**ANALYSIS OF ALGORITHMS AND PROBLEM COMPLEXITY**]: Numerical Algorithms and Problems; G.3 [**PROBABILITY AND STATISTICS**]; G.4 [**MATHEMATICAL SOFTWARE**]; E.2 [**DATA STORAGE REPRESENTATIONS**]: Hash-table representations

---
[*]This work was done while visiting Carnegie Mellon University. It is supported by the Danish National Research Foundation under the Sapere Aude program.

## 1. INTRODUCTION
Several computational problems can be phrased in terms of matrix products where the normalized result is expected to consist mostly of numerically small entries:

- Given $m$ samples of a multivariate random variable $(X_1, \dots, X_n)$, compute the *sample covariance matrix* which is used in statistical analyses. If most pairs of random variables are independent, the corresponding entries of the sample covariance matrix will be concentrated around zero.

- Linearly transform all column vectors in a matrix $B$ to an orthogonal basis $A^T$ in which the columns of $B$ are approximately sparse. Such batch transformations are common in lossy data compression algorithms such as JPEG, using properties of specific orthogonal bases to facilitate fast computation.

In both cases, an approximation of the matrix product may be as good as an exact computation, since the main issue is to identify large entries in the result matrix.

In this paper we consider $n$-by-$n$ matrices with real values, and devise a combinatorial algorithm for the special case of computing a matrix product $AB$ that is "compressible". For example, if $AB$ is sparse it falls into our class of compressible products, and we are able to give an efficient algorithm. More generally, if the Frobenius norm of $AB$ is dominated by a sparse subset of the entries, we are able to quickly compute a good approximation of the product $AB$. Our method can be seen as a compressed sensing method for the matrix product, with the nonstandard idea that the sketch of $AB$ is computed without explicitly constructing $AB$. The main technical idea is to use FFT [11] to efficiently compute a linear sketch of an outer product of two vectors. We also make use of error-correcting codes in a novel way to achieve recovery of the entries of $AB$ having highest magnitude in near-linear time.

Our main conceptual messages are:

- It is possible to derive a fast and simple, "combinatorial" algorithm for matrix multiplication in two cases: When the output matrix is sparse, and when an additive error on each output entry is acceptable.

- It is interesting to consider the use of compressed sensing techniques for computational problems where results (or important intermediate results) can be represented using sparse vectors. We outline some such targets in the conclusion.

- The interface between theoretical computer science and statistics is an area where there is high potential for cross-fertilization (see also [21] for arguments in this direction).

## 1.1 Related work

*Matrix multiplication with sparse output.*

Lingas [20] considered the problem of computing a matrix product $AB$ with at most $\bar{b}$ entries that are not *trivially* zero. A matrix entry is said to be trivially zero if every term in the corresponding dot product is zero. In general $\bar{b}$ can be much larger than the number $b$ of nonzeros because zeros in the matrix product may be due to cancellations. Lingas showed, by a reduction to fast rectangular matrix multiplication, that this is possible in time $\mathcal{O}(n^2 \bar{b}^{0.188})$. Observe that for $\bar{b} = n^2$ this becomes identical to the $\mathcal{O}(n^{2.376})$ bound by Coppersmith and Winograd [12].

Yuster and Zwick [25] devised asymptotically fast algorithms for the case of sparse *input* matrices, using a matrix partitioning idea. Amossen and Pagh [4] extended this result to be more efficient in the case where also the output matrix is sparse. In the dense input setting of Lingas, this leads to an improved time complexity of $\mathcal{O}(n^{1.724} \bar{b}^{0.408})$ for $n \leq \bar{b} \leq n^{1.25}$.

Iwen and Spencer [19] showed how to use compressed sensing to compute a matrix product $AB$ in time $\mathcal{O}(n^{2+\varepsilon})$, for any given constant $\varepsilon > 0$, in the special case where *each column* of $AB$ contains at most $n^{0.29462}$ nonzero values. (Of course, by symmetry the same result holds when there is sparseness in each row.)

All the results described above work by reduction to fast rectangular matrix multiplication, so the algorithms are not "combinatorial." However, Lingas [20] observed that a time complexity of $\mathcal{O}(n^2 + \bar{b}n)$ is achieved by the column-row method, a simple combinatorial algorithm. Also, replacing the fast rectangular matrix multiplication in the result of Iwen and Spencer [19] by a naïve matrix multiplication algorithm, and making use of randomized sparse recovery methods (see [15]), leads to a combinatorial algorithm running in time $\tilde{\mathcal{O}}(n^2 + nb)$ when each column of $AB$ has $\mathcal{O}(b/n)$ nonzero values.

*Approximate matrix multiplication.*

The result of [19] is not restricted to sparse matrix products: Their algorithm is shown to compute an approximate matrix product in time $\mathcal{O}(n^{2+\varepsilon})$ assuming that the result can be approximated well by a matrix with sparse column vectors. The approximation produced is one with at most $n^{0.29462}$ nonzero values in each column, and is almost as good as the best approximation of this form. However, if some column of $AB$ is dense, the approximation may differ significantly from $AB$.

Historically, Cohen and Lewis [10] were the first to consider randomized algorithms for approximate matrix multiplication, with theoretical results restricted to the case where input matrices do not have negative entries. Suppose $A$ has column vectors $a_1, \ldots, a_n$ and $B$ has row vectors $b_1, \ldots, b_n$. The product of $A$ and $B$ can be written as a sum of $n$ outer products:

$$AB = \sum_{k=1}^{n} a_k b_k \ . \tag{1}$$

The method of Cohen and Lewis can be understood as sampling each outer product according to the weight of its entries, and combining these samples to produce a matrix $C$ where each entry is an unbiased estimator for $(AB)_{ij}$. If $n^2 c$ samples are taken, for a parameter $c \geq 1$, the difference between $C$ and $AB$ can be bounded, whp., in terms of the Frobenius norm[1] of $AB$, namely

$$||AB - C||_F = \mathcal{O}(||AB||_F/\sqrt{c}) \ .$$

(This is not shown in [10], but follows from the fact that each estimator has a scaled binomial distribution.)

Drineas, Kannan, and Mahoney [14] showed how a simpler sampling strategy can lead to a good approximation of the form $CR$, where matrices $C$ and $R$ consist of $c$ columns and $c$ rows of $A$ and $B$, respectively. Their main error bound is in terms of the Frobenius norm of the difference: $||AB - CR||_F = \mathcal{O}(||A||_F ||B||_F/\sqrt{c})$. The time to compute $CR$ using the classical algorithm is $\mathcal{O}(n^2 c)$ — asymptotically faster results are possible by fast rectangular matrix multiplication. Drineas et al. also give bounds on the elementwise differences $|(AB-CR)_{ij}|$, but the best such bound obtained is of size $\Omega(M^2 n/\sqrt{c})$, where $M$ is the magnitude of the largest entry in $A$ and $B$. This is a rather weak bound in general, since the largest possible magnitude of an entry in $AB$ is $M^2 n$.

Sarlós [23] showed how to achieve the same Frobenius norm error guarantee using $c$ AMS sketches [2] on rows of $A$ and columns of $B$. Again, if the classical matrix multiplication algorithm is used to combine the sketches, the time complexity is $\mathcal{O}(n^2 c)$. This method gives a stronger error bound for each individual entry of the approximation matrix. If we write an entry of $AB$ as a dot product, $(AB)_{ij} = \tilde{a}_i \cdot \tilde{b}_j$, the magnitude of the additive error is $\mathcal{O}(||\tilde{a}_i||_2 ||\tilde{b}_j||_2/\sqrt{c})$ with high probability (see [23, 1]). In contrast to the previous results, this approximation can be computed in a single pass over the input matrices. Clarkson and Woodruff [8] further refine the results of Sarlós, and show that the space usage is nearly optimal in a streaming setting.

## 1.2 New results

In this paper we improve existing results in cases where the matrix product is "compressible" — in fact, we produce a compressed representation of $AB$. Let $N \leq 2n^2$ denote the number of nonzero entries in $A$ and $B$. We obtain an approximation $C$ by a combinatorial algorithm running in time $\tilde{\mathcal{O}}(N + nb)$, making a single pass over the input while using space $\mathcal{O}(b \lg n)$, such that:

- If $AB$ has at most $b$ nonzero entries, $C = AB$ whp.

- If $AB$ has Frobenius norm $q$ when removing its $b$ largest entries, the error of each entry is bounded, whp., by

$$|C_{ij} - (AB)_{ij}| < q/\sqrt{b} \ .$$

Compared to Cohen and Lewis [10] we avoid the restriction that input matrices cannot contain negative entries. Also, their method will produce only an approximate result

---

[1]The *Frobenius norm* of a matrix $A$ is defined as

$$||A||_F = \sqrt{\sum_{i,j} A_{ij}^2} \ .$$

even when $AB$ is sparse. Finally, their method inherently uses space $\Theta(n^2)$, and hence is not able to exploit compressibility to achieve smaller space usage.

Our algorithm is faster than existing matrix multiplication algorithms for sparse outputs [4, 20] whenever $b < n^{6/5}$, as well as in situations where a large number of cancellations mean that $b \ll \bar{b}$. As a more conceptual contribution it is to our knowledge the only "simple" algorithm to significantly improve on Strassen's algorithm for sparse outputs with many entries that are not trivially zero.

The simple, combinatorial algorithms derived from Drineas et al. [14] and Sarlós [23] yield error guarantees that are generally incomparable with those achieved here, when allowing same time bound, i.e., $c = \Theta(b/n)$. The Frobenius error bound we achieve is:

$$||AB - C||_F \leq ||AB||_F \sqrt{n/c} \ .$$

Our result bears some similarity to the result of Iwen and Spencer [19], since both results be seen as compressed sensing of the product $AB$. One basic difference is that Iwen and Spencer perform compressed sensing on each column of $AB$ ($n$ sparse signals), while we treat the whole matrix $AB$ as a single sparse signal. This means that we are robust towards skewed distribution of large values among the columns.

## 2. ALGORITHM AND ANALYSIS

We will view the matrix $AB$ as the set of pairs $(i, j)$, where the weight of item $(i, j)$ is $(AB)_{ij}$. Our approach is to compute a linear sketch $p_{a_k b_k}$ for each outer product of (1), and then compute $\sum_k p_{a_k b_k}$ to obtain a sketch for $AB$. For exposition, we first describe how to compute an AMS sketch of this weighted set in time $\mathcal{O}(n)$, and then extend this to the more accurate COUNT SKETCH. In the following we use $[n]$ to denote the set $\{1, \ldots, n\}$.

### 2.1 AMS sketches

Alon, Matias, and Szegedy [2] described and analyzed the following approach to sketching a data stream $z_1, z_2, \ldots$, where item $z_i \in [n]$ has weight $w_i$: Take a 2-wise independent[2] function $s : [n] \rightarrow \{-1, +1\}$ (which we will call the *sign function*), and compute the sum $X = \sum_i s(z_i) w_i$ (which we refer to as the *AMS sketch*). We will use a sign function on pairs $(i, j)$ that is a product of sign functions on the coordinates: $s(i, j) = s_1(i) s_2(j)$. Indyk and McGregor [18], and Braverman et al. [5] have previously analyzed moments of AMS sketches with hash functions of this form. However, for our purposes it suffices to observe that $s(i, j)$ is 2-wise independent if $s_1$ and $s_2$ are 2-wise independent. The AMS sketch of an outer product $uv$, where by definition $(uv)_{ij} = u_i v_j$, is:

$$\sum_{(i,j) \in [n] \times [n]} s(i,j) (uv)_{ij} = \left( \sum_{i=1}^n s_1(i) u_i \right) \left( \sum_{j=1}^n s_2(j) v_j \right) \ .$$

That is, the sketch for the outer product is simply the product of the sketches of the two vectors (using different hash functions). A single AMS sketch has variance roughly $||uv||_F^2$, which is too large for our purposes. Taking the average of $b$ such sketches to reduce the variance by a factor $\sqrt{b}$ would increase the time to retrieve an estimator for an entry in the matrix product by a factor of $b$. By using a different sketch we can avoid this problem, and additionally get better estimates for compressible matrices.

---

[2]We will use "$h$ is a a $k$-wise independent hash function" as a shorthand for "$h$ is chosen uniformly at random from a $k$-wise independent family of hash functions, independently of all other random choices made by the algorithm".

### 2.2 Count sketches

Our algorithm will use the COUNT SKETCH of Charikar, Chen and Farach-Colton [7], which has precision at least as good as the estimator obtained by taking the average of $b$ AMS sketches, but is much better for skewed distributions. The method maintains a sketch of any desired size $b$, using a 2-wise independent *splitting* function $h : [n] \rightarrow \{0, \ldots, b-1\}$ to divide the items into $b$ groups. For each group, an AMS sketch is maintained using a 2-wise independent hash function $s : [n] \rightarrow \{-1, +1\}$. That is, the sketch is the vector $(c_0, \ldots, c_{b-1})$ where $c_k = \sum_{i, h(z_i)=k} s(z_i) w_i$. An unbiased estimator for the total weight of an item $z$ is $c_{h(z)} s(z)$. To obtain stronger guarantees, one can take the median of several estimators constructed as above (with different sets of hash functions) — we return to this in section 3.3.

*Sketching a matrix product naïvely.*

Since COUNT SKETCH is linear we can compute the sketch for $AB$ by sketching each of the outer products in (1), and adding the sketch vectors. Each outer product has $\mathcal{O}(n^2)$ terms, which means that a direct approach to computing its sketch has time complexity $\mathcal{O}(n^2)$, resulting in a total time complexity of $\mathcal{O}(n^3)$.

*Improving the complexity*

We now show how to improve the complexity of the outer product sketching from $\mathcal{O}(n^2)$ to $\mathcal{O}(n + b \lg b)$ by choosing the hash functions used by COUNTSKETCH in a "decomposable" way, and applying FFT. We use the sign function $s(i, j)$ defined in section 2.1, and similarly decompose the function $h$ as follows: $h(i, j) = h_1(i) + h_2(j) \bmod b$, where $h_1$ and $h_2$ are chosen independently at random from a 3-wise independent family. It is well-known that this also makes $h$ 3-wise independent [6, 22]. Given a vector $u \in \mathbf{R}^n$ and functions $h_t : [n] \rightarrow \{0, \ldots, b-1\}$, $s_t : [n] \rightarrow \{-1, +1\}$ we define the following polynomial:

$$p_u^{h_t, s_t}(x) = \sum_{i=1}^n s_t(i) u_i x^{h_t(i)} \ .$$

The polynomial can be represented either in the standard basis as the vector of coefficients of monomials $x^0, \ldots, x^{b-1}$, or in the *discrete Fourier basis* as the vector

$$(p_u^{h_t, s_t}(\omega^0), p_u^{h_t, s_t}(\omega^1), \ldots, p_u^{h_t, s_t}(\omega^{b-1})),$$

where $\omega$ is a complex number such that $\omega^b = 1$. The efficient transformation to the latter representation is known as the fast Fourier transform (FFT), and can be computed in time $\mathcal{O}(b \log b)$ when $b$ is a power of 2 [11]. Taking componentwise products of the vectors representing $p_u^{h_1, s_1}$ and $p_v^{h_2, s_2}$ in the Fourier basis we get a vector $p^*$ where $p_t^* = p_u^{h_1, s_1}(\omega^t) p_v^{h_2, s_2}(\omega^t)$. Now consider the following polynomial:

$$p_{uv}^*(x) = \sum_k \sum_{\substack{i,j \\ h(i,j)=k}} s(i,j) (uv)_{ij} x^k$$

Using $\omega^{t\,h(i,j)} = \omega^{t\,h_1(i)+t\,h_2(j)}$ we have that

$$p_{uv}^*(\omega^t) = \sum_{i,j} s(i,j)\,(uv)_{ij}\,\omega^{h(i,j)t}$$
$$= \left(\sum_i s_1(i)\,u_i\,\omega^{t\,h_1(i)}\right)\left(\sum_j s_2(j)\,v_j\,\omega^{t\,h_2(j)}\right)$$
$$= p_t^* \ .$$

That is, $p^*$ is the representation, in the discrete Fourier basis, of $p_{uv}^*$. Now observe that the coefficients of $p_{uv}^*(x)$ are the entries of a COUNT SKETCH for the outer product $uv$ using the sign function $s(i,j)$ and splitting function $h(i,j)$. Thus, applying the inverse FFT to $p^*$ we compute the COUNT SKETCH of $uv$.

The pseudocode of figure 1, called with parameter $d = 1$, summarizes the encoding and decoding functions discussed so far. For simplicity the pseudocode assumes that the hash functions involved are fully random. A practical implementation of the involved hash functions is character-based tabulation [22], but for the best theoretical space bounds we use polynomial hash functions [13].

*Time and space analysis.*

We analyze each iteration of the outer loop. Computing $p_{uv}(x)$ takes time $\mathcal{O}(n + b \lg b)$, where the first term is the time to construct the polynomials, and the last term is the time to multiply the polynomials, using FFT [11]. Computing the sketch for each outer product and summing it up takes time $\mathcal{O}(n^2 + nb \lg b)$. Finally, in time $\mathcal{O}(n^2)$ we can obtain the estimate for each entry in $AB$.

The analysis can be tightened when $A$ and $B$ are sparse or rectangular, and it suffices to compute the sketch that allows random access to the estimate $C$. Suppose that $A$ is $n_1$-by-$n_2$, and $B$ is $n_2$-by-$n_3$, and they contain $N \ll n^2$ nonzero entries. It is straightforward to see that each iteration runs in time $\mathcal{O}(N + n_2 b \lg b)$, assuming that $A$ and $B$ are given in a form that allows the nonzero entries of a column (row) to be retrieved in linear time.

The required hash functions, with constant evaluation time, can be stored in space $\mathcal{O}(d)$ using polynomial hash functions [13]. The space required for the rest of the computation is $\mathcal{O}(db)$, since we are handling $\mathcal{O}(d)$ polynomials of degree less than $b$. Further, access to the input is restricted to a single pass if $A$ is stored in column-major order, and $B$ is stored in row-major order.

LEMMA 1. COMPRESSEDPRODUCT$(A, B, b, d)$ *runs in time* $\mathcal{O}(d(N + n_2 b \lg b))$, *and uses space for* $\mathcal{O}(db)$ *real numbers, in addition to the input.*

We note that the time bound may be much smaller than $n^2$, which is the number of entries in the approximate product $C$. This is because $C$ is not constructed explicitly. In section 4 we address how to efficiently extract the $b$ largest entries of $C$.

## 3. ERROR ANALYSIS

We can obtain two kinds of guarantees on the approximation: One in terms of the Frobenius norm (section 3.1), which applies even if we use just a single set of hash functions ($d = 1$), and stronger guarantees (section 3.3) that require the use of $d = \mathcal{O}(\lg n)$ hash functions. Section 3.2

```
function COMPRESSEDPRODUCT(A, B, b, d)
    for t := 1 to d do
        s_1[t], s_2[t] ∈_R Maps({1,...,n} → {-1,+1})
        h_1[t], h_2[t] ∈_R Maps({1,...,n} → {0,...,b-1})
        p[t] := 0
        for k := 1 to n do
            (pa, pb) := (0, 0)
            for i := 1 to n do
                pa[h_1(i)] := pa[h_1(i)] + s_1[t](i) A_{ik}
            end for
            for j := 1 to n do
                pa[h_2(j)] := pb[h_2(j)] + s_2[t](j) B_{kj}
            end for
            (pa, pb) := (FFT(pa), FFT(pb))
            for z := 1 to b do
                p[t][z] := p[t][z] + pa[z] pb[z]
            end for
        end for
    end for
    for t := 1 to d do p[t] := FFT^{-1}(p[t]) end for
    return (p, s_1, s_2, h_1, h_2)
end

function DECOMPRESS(i, j)
    for t := 1 to d do
        X_t := s_1[t](i) s_2[t](j) p[t][(h_1(i) + h_2(j)) mod b]
    return MEDIAN(X_1,...,X_d)
end
```

Figure 1: Method for encoding an approximate representation of $AB$ of size $bd$, and corresponding method for decoding its entries. Maps$(D \to C)$ denotes the set of functions from $D$ to $C$, and $\in_R$ denotes independent, random choice. (Limited randomness is sufficient to obtain our guarantees, but this is not reflected in the pseudocode.) We use $\mathbf{0}$ to denote a zero vector (or array) of suitable size that is able to hold complex numbers. FFT$(p)$ denotes the discrete fourier transform of vector $p$, and FFT$^{-1}$ its inverse.

considers the application of our result to covariance matrix estimation.

### 3.1 Frobenius norm guarantee

THEOREM 2. *For $d = 1$ and any $(i^*, j^*)$, the function call* DECOMPRESS$(i^*, j^*)$ *computes an unbiased estimator for* $(AB)_{i^*j^*}$ *with variance bounded by* $\|AB\|_F^2 / b$.

PROOF. For $i, j \in \{1, \ldots, n\}$ let $K_{i,j}$ be the indicator variable for the event $h(i,j) = h(i^*, j^*)$. Since $h$ is 3-wise independent these events are 2-wise independent. We can write $X$ as:

$$X = s(i^*, j^*) \sum_{i,j} K_{i,j} s(i,j) (AB)_{ij} \ .$$

Observe that $K(i^*, j^*) = 1$, $\mathbf{E}[s(i^*, j^*) s(i, j)] = 0$ whenever $(i, j) \neq (i^*, j^*)$, and $\mathbf{E}[s(i^*, j^*)^2] = 1$. This implies that $\mathbf{E}[X] = (AB)_{i^*j^*}$. To bound the variance of $X$ we rewrite

it as:

$$X = (AB)_{i^*j^*} + s(i^*,j^*) \sum_{(i,j) \neq (i^*,j^*)} K_{i,j} s(i,j)(AB)_{ij} \quad (2)$$

Since $s(i^*,j^*)^2 = 1$ and the values $K_{i,j}$, $i,j \in \{1,\ldots,n\}$, and $s(i,j)$, $i,j \in \{1,\ldots,n\}$ are 2-wise independent the terms have covariance zero, so the variance is simply the sum

$$\sum_{(i,j) \neq (i^*,j^*)} \mathbf{Var}(K_{i,j} s(i,j)(AB)_{ij}) \ .$$

We have that $\mathbf{E}[K_{i,j} s(i,j)(AB)_{ij}] = 0$, so the variance of each term is equal to its second moment:

$$\mathbf{E}[(K_{i,j} s(i,j)(AB)_{ij})^2] = (AB)_{ij}^2 \mathbf{E}[K_{i,j}] = (AB)_{ij}^2/b \ .$$

Summing over all terms we get that $\mathbf{Var}(X) \leq ||AB||_F^2/b$. □

As a consequence of the lemma, we get from Chebychev's inequality that each estimate is accurate with probability 3/4 up to an additive error of $2||AB||_F/\sqrt{b}$. For sufficiently large $d = \mathcal{O}(\lg n)$ this holds for all entries with high probability, by Chernoff bounds.

## 3.2 Covariance matrix estimation

We now consider the application of our result to covariance matrix estimation from a set of samples. The covariance matrix captures pairwise correlations among the components of a multivariate random variable $X = (X_1,\ldots,X_n)^T$. It is defined as $\text{cov}(X) = \mathbf{E}[(X-\mathbf{E}[X])(X-\mathbf{E}[X])^T]$. We can arrange observations $x_1,\ldots,x_m$ of $X$ as columns in an $n$-by-$m$ matrix $A$. Figure 2 illustrates how approximate matrix multiplication can be used to find correlations among rows of $A$ (corresponding to components of $X$). In the following we present a theoretical analysis of this approach.

The *sample mean* of $X$ is $\bar{x} = \frac{1}{m}\sum_{i=1}^m x_i$. The *sample covariance matrix* $Q$ is an unbiased estimator of $\text{cov}(X)$, given by:

$$Q = \tfrac{1}{m-1} \sum_{i=1}^m (x_i - \bar{x})(x_i - \bar{x})^T \ .$$

Let $\bar{x}1$ denote the $n$-by-$m$ matrix that has all columns equal to $\bar{x}$. Then we can write $Q$ as a matrix product:

$$Q = \tfrac{1}{m-1}(A - \bar{x}1)(A - \bar{x}1)^T \ .$$

To simplify calculations we consider computation of $\tilde{Q}$ which is derived from $Q$ by setting entries on the diagonal to zero. Notice that a linear sketch of $Q$ can be transformed easily into a linear sketch of $\tilde{Q}$, and that a sketch of $\tilde{Q}$ also allows us to quickly approximate $Q$. Entry $\tilde{Q}_{ij}$, $i \neq j$ is a random variable that has expectation 0 if $X_i$ and $X_j$ are independent. It is computed as $\frac{1}{m-1}$ times the sum over $m$ observations of $(X_i - \mathbf{E}[X_i])(X_j - \mathbf{E}[X_j])$. Assuming independence and using the formula from [17], this means that its variance is $\frac{m}{(m-1)^2}\text{Var}(X_i)\text{Var}(X_j) < \frac{4}{m}\text{Var}(X_i)\text{Var}(X_j)$, for $m \geq 2$. If $\text{cov}(X)$ is a diagonal matrix (i.e., every pair of variables is independent), the expected squared Frobenius norm of $\tilde{Q}$ is:

$$\mathbf{E}[||\tilde{Q}||_F^2] = 2\sum_{i<j} \mathbf{E}[Q_{ij}^2] = 2\sum_{i<j} \text{Var}(Q_{ij})$$

$$< \tfrac{8}{m} \sum_{i<j} \text{Var}(X_i)\text{Var}(X_j)$$

$$< \tfrac{4}{m}\left(\sum_i \text{Var}(X_i)\right)^2 \ .$$

In a statistical test for pairwise independence one will assume independence, and test if the sample covariance matrix is indeed close to diagonal. We can derive an approximation guarantee from Theorem 2 for the sketch of $\tilde{Q}$ (and hence $Q$), assuming the hypothesis that $\text{cov}(X)$ is diagonal. If this is not true, our algorithm will still be computing an unbiased estimate of $\text{cov}(X)$, but the observed variance in each entry will be larger.

THEOREM 3. *Consider $m$ observations of random variables $X_1,\ldots,X_n$ that are pairwise independent. We can compute in time $\tilde{\mathcal{O}}((n+b)m)$ and space $\tilde{\mathcal{O}}(b)$ an unbiased approximation to the sample covariance matrix with additive error on each entry (whp.) $\mathcal{O}(\sum_{i=1}^n \text{Var}(X_i)/\sqrt{mb})$.*

No similar result follows from the method of Cohen and Lewis [10], which gives no theoretical guarantees when applied to matrices with negative entries. Similarly, the algorithms of Drineas et al. [14] and Sarlós [23] do not have sufficiently strong guarantees on the error of single entries to imply Theorem 3. We note that a similar result could be obtained by the method of Iwen and Spencer [19].

The special case of indicator random variables is of particular interest. Then $\sum_{i=1}^n \text{Var}(X_i) \leq n$, and if $m \geq n$ we can achieve additive error $o(1)$ in time slightly superlinear in the size of the input matrix:

COROLLARY 4. *Consider $m$ observations of indicator random variables $X_1,\ldots,X_n$ that are pairwise independent. Using space $b \geq n$ and time $\tilde{\mathcal{O}}(mb)$ we can compute an approximation to the sample covariance matrix with additive error $\mathcal{O}(n/\sqrt{mb})$.*

## 3.3 Tail guarantees

We now provide a stronger analysis of our matrix multiplication algorithm, focusing on the setting in which we compute $d = \mathcal{O}(\lg n)$ independent sketches as described in section 2.2, and the estimator returned is the median.

### 3.3.1 Sparse outputs.

We first show that sparse outputs are computed exactly with high probability.

THEOREM 5. *Suppose $AB$ has at most $b/8$ nonzero entries, and $d \geq 6\lg n$. Then COMPRESSEDPRODUCT($A,B,b,d$) together with DECOMPRESS correctly computes $AB$ with probability $1 - o(1)$.*

PROOF. Let $S_0$ denote the set of coordinates of nonzero entries in $AB$. Consider again the estimator (2) for $(AB)_{i^*j^*}$. We observe that $X \neq (AB)_{i^*j^*}$ can only happen when $K_{i,j} \neq 0$ for some $(i,j) \neq (i^*,j^*)$ with $(AB)_{ij} \neq 0$, i.e., $(i,j) \in S_0$ and $h(i,j) = h(i^*,j^*)$. Since $h$ is 2-wise independent with range $b$ and $|S_0| \leq b/8$, this happens with probability at most $1/8$. The expected number of sketches with $X \neq (AB)_{i^*j^*}$ is therefore at most $d/8$. If less than $d/2$ of the sketches have $X \neq (AB)_{i^*j^*}$, the median will be $(AB)_{i^*j^*}$. By Chernoff bounds, the probability that the expected value $d/8$ is exceeded by a factor 4 is $(e^{4-1}/4^4)^{d/10} = o(2^{-d/3})$. In particular, if we choose $d = 6\lg n$ then the probability that the output is correct in *all* entries is $1 - o(1)$. □

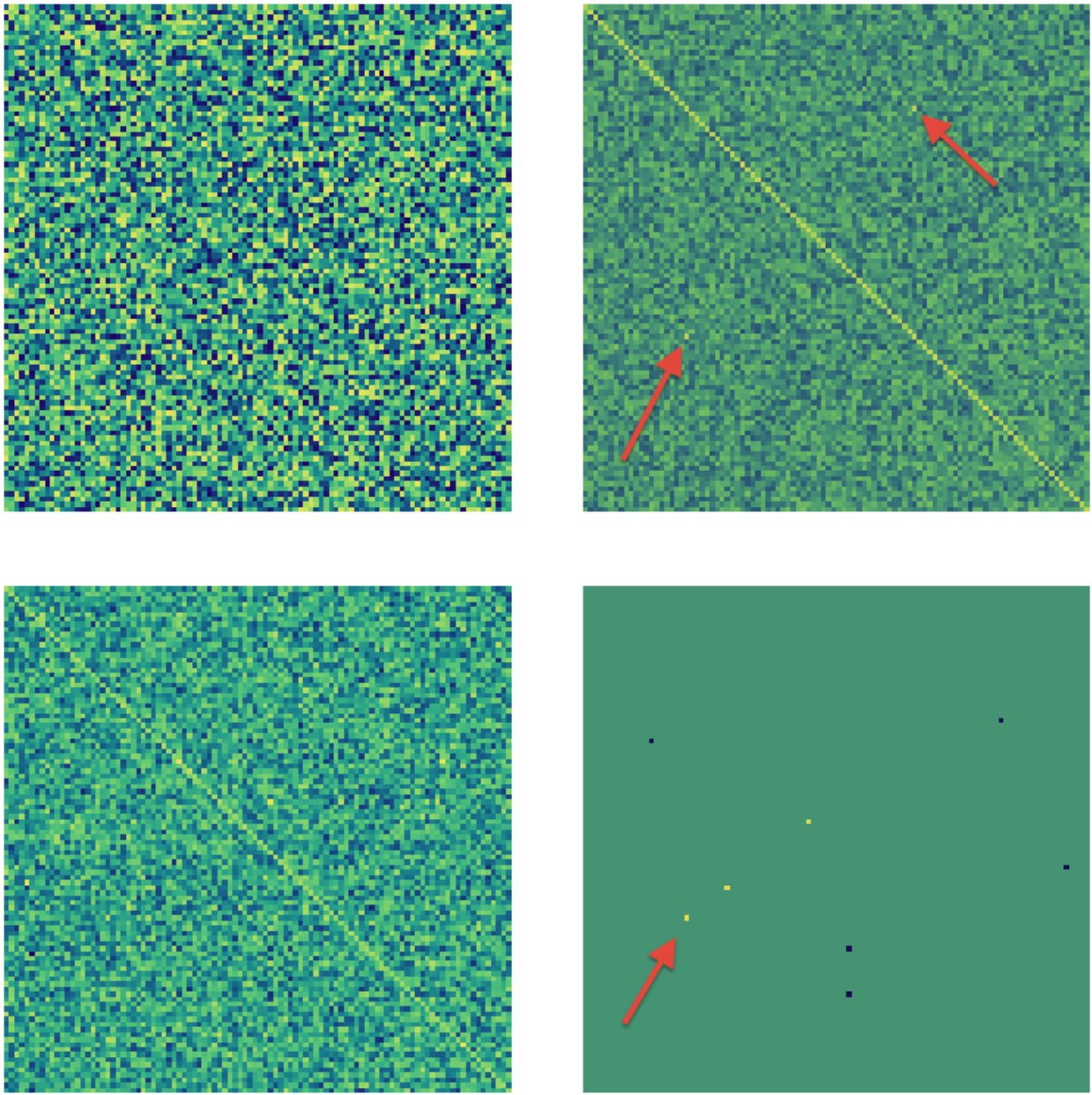

**Figure 2:** Finding correlations using approximate matrix multiplication. *Upper left:* 100-by-100 random matrix $A$ where all entries are sampled from $[-1;1]$, independently except rows **21** and **66** which are positively correlated. *Upper right:* $AA^T$ has mostly numerically small values off diagonals, expect entries **(66,21)** and **(21,66)** corresponding to the correlated rows in $A$. *Lower left:* Approximation of $AA^T$ output by our algorithm using $b = 2000$. *Lower right:* After subtracting the contribution of diagonal elements of $AA^T$ and thresholding the resulting approximation, a small set of entries remains that are "candidates for having a large value", including **(66,21)**.

### 3.3.2 Skewed distributions.

Our next goal is to obtain stronger error guarantees in the case where the distribution of values in $AB$ is skewed such that the Frobenius norm is dominated by the $b/20$ largest entries.

Let $\mathrm{Err}_F^k(M)$ denote the squared Frobenius norm (i.e., sum of entries squared) of a matrix that is identical to matrix $M$ except for its $k$ largest entries (absolute value), where it is zero.

THEOREM 6. *Suppose that* $d \geq 6 \lg n$. *Then* DECOMPRESS *in conjunction with* COMPRESSEDPRODUCT$(A, B, b, d)$ *computes a matrix $C$ such that for each entry $C_{ij}$ we have*

$$|C_{ij} - (AB)_{ij}| < 12\sqrt{\mathrm{Err}_F^{b/20}(AB)/b}$$

*with probability* $1 - o(n^{-2})$.

PROOF. Consider again equation (2) that describes a single estimator for $(AB)_{i^*j^*}$. Let $\mathbf{v}$ be the length $n^2 - 1$ vector with entries $(K_{i,j}(AB)_{ij})_{ij}$, ranging over all $(i,j) \neq (i^*, j^*)$. The error of $X$ is $s(i^*, j^*)$ times the dot product of $\mathbf{v}$ and the $\pm 1$ vector represented by $s$. Since $s$ is 2-wise independent, $\mathbf{Var}(X) \leq \|\mathbf{v}\|_2^2$. Let $L$ denote the set of coordinates of the $b/20$ largest entries in $AB$ (absolute value), different from $(i^*, j^*)$, with ties resolved arbitrarily. We would like to argue that with probability $9/10$ two events hold simultaneously:

- $K(i,j) = 0$ for all $(i,j) \in L$.
- $\|\mathbf{v}\|_2^2 \leq \sigma^2$, where $\sigma^2 = 20\,\mathrm{Err}_F^{b/20}(AB)/b$.

When this is the case we say that $\mathbf{v}$ is "good". The first event holds with probability $19/20$ by the union bound, since $\Pr[K(i,j) = 1] = 1/b$. To analyze the second event we focus on the vector $\mathbf{v}'$ obtained from $\mathbf{v}$ by fixing $K(i,j) = 0$ for $(i,j) \in L$. The expected value of $\|\mathbf{v}'\|_2^2$ is bounded by $\sigma^2/20$. Thus by Markov's inequality we have $\|\mathbf{v}'\|_2^2 \leq \sigma^2$ with probability $19/20$. Note that when the first event holds we have $\mathbf{v}' = \mathbf{v}$. So we conclude that $\mathbf{v}$ is good with probability at least $9/10$.

For each estimator $X$, since $\mathbf{Var}(X) \leq \|\mathbf{v}\|_2^2$ the probability that the error is of magnitude $t$ or more is at most $\|\mathbf{v}\|_2^2/t^2$ by Chebychev's inequality. So for a good vector $\mathbf{v}$ the probability that $t^2 \geq 7\sigma^2 \geq 7\|\mathbf{v}\|_2^2$ is at most $1/49$. Thus for each estimator the probability that it is based on a good vector and has error less than

$$\sqrt{7\sigma^2} \leq 12\sqrt{\mathrm{Err}_F^{b/20}(AB)/b}$$

is at least $1 - 1/10 - 1/49 > 7/8$.

Finally observe that it is unlikely that $d/2$ or more estimators have larger error. As in the proof of Theorem 5 we get that the probability that this occurs is $o(2^{-d/3}) = o(n^{-2})$. Thus, the probability that the median estimator has larger error is $o(n^{-2})$. □

## 4. SUBLINEAR RESULT EXTRACTION

In analogy with the sparse recovery literature (see [15] for an overview) we now consider the task of extracting the most significant coefficients of the approximation matrix $C$ in time $o(n^2)$. In fact, if we allow the compression algorithm to use a factor $\mathcal{O}(\lg n)$ more time and space, the time complexity for decompression will be $\mathcal{O}(b\lg^2 n)$. Our main tool is error-correcting codes, previously applied to the sparse recovery problem by Gilbert et al. [16]. However, compared to [16] we are able to proceed in a more direct way that avoids iterative decoding. We note that a similar result could be obtained by a 2-dimensional dyadic decomposition of $[n] \times [n]$, but it seems that this would result in time $\mathcal{O}(b\lg^3 n)$ for decompression.

For $\Delta \geq 0$ let $S_\Delta = \{(i,j) \mid |(AB)_{ij}| > \Delta\}$ denote the set of entries in $AB$ with magnitude larger than $\Delta$, and let $L$ denote the $b/\kappa$ largest entries in $AB$, for some constant $\kappa$. Our goal is to compute a set $S$ of $\mathcal{O}(b)$ entries that contains $L \cap S_\Delta$ where $\Delta = \mathcal{O}(\sqrt{\mathrm{Err}_F^{b/\kappa}(AB)/b})$. Intuitively, with high probability we should output entries in $L$ if their magnitude is significantly above the standard deviation of entries of the approximation $C$.

### 4.1 Approach

The basic approach is to compute a sequence of COUNT SKETCHES using the *same set of hash functions* to approximate different submatrices of $AB$. The set of sketches that contain a particular entry will then reveal (with high probability) the location of the entry. The submatrices are constructed using a good error-correcting code

$$E : [n] \to \{0,1\}^\ell,$$

where $\ell = \mathcal{O}(\lg n)$. Let $I_{E_r}$ denote the diagonal matrix where entry $(i,i)$ equals $E(i)_r$, bit number $r$ of $E(i)$. Then $I_{E_r}A$ is the matrix that is derived from $A$ by changing entries to zero in those rows $i$ for which $E(i)_r = 0$. Similarly, we can derive $BI_{E_r}$ from $B$ by changing entries to zero in columns $j$ where $E(j)_r = 0$. The matrix sketches that we compute are:

$$\begin{aligned} C_{r\cdot} &= (I_{E_r}A)B, \text{ for } r \in \{1,\ldots,\ell\}, \text{ and} \\ C_{\cdot r} &= A(BI_{E_{r-\ell}}), \text{ for } r \in \{\ell+1,\ldots,2\ell\} \end{aligned} \quad (3)$$

We aim to show the following result.

THEOREM 7. *Assume $d = \mathcal{O}(\lg n)$ is sufficiently large. There exists a constant $\kappa$ such that if $\Delta \geq \kappa\sqrt{\mathrm{Err}_F^{b/\kappa}(AB)/b}$ then* FINDSIGNIFICANTENTRIES$(\Delta)$ *returns a set of $\mathcal{O}(b)$ positions that includes the positions of the $b/\kappa$ entries in $AB$ having the highest magnitudes, possibly omitting entries with magnitude below $\Delta$. The running time is $\mathcal{O}(b\lg^2 n)$, space usage is $\mathcal{O}(b\lg n)$, and the result is correct with probability $1 - \frac{1}{n}$.*

Combining this with Theorem 5 and Lemma 1 we obtain:

COROLLARY 8. *Let $A$ be an $n_1$-by-$n_2$ matrix, and $B$ an $n_2$-by-$n_3$ matrix, with $N$ nonzero entries in total. Further suppose that $AB$ is known to have at most $b$ nonzero entries. Then a sparse representation of $AB$, correct with probability $1 - \frac{1}{n}$, can be computed in time $\mathcal{O}(N + n_2 b \lg b + b \lg^2 n)$, using space $\mathcal{O}(b \lg n)$ in addition to the input.*

```
function FINDSIGNIFICANTENTRIES(Δ)
  S := ∅
  for t := 1 to d do
    for k := 1 to b do
      for r := 1 to 2ℓ do X_r := |p^(t,r)[k]| end for
      s := ε
      for r := 1 to 2ℓ do
        if X_r > Δ/2 then s := s||1 else s := s||0
      end for
      (i,j) := DECODE(s)
      INSERT((i,j), S)
    end for
  end for
  for (i,j) ∈ S do
    if |{(i,j)} ∩ S| < d/2 then DELETE((i,j), S) end if
  end for
  return S
end
```

**Figure 3:** Method for computing the positions of $\mathcal{O}(b)$ significant matrix entries of magnitude $\Delta$ or more. String concatenation is denoted $||$, and $\epsilon$ denotes the empty string. DECODE($s$) decodes the (corrupted) codewords formed by the bit string $s$ (which must have length a multiple of $\ell$), returning an arbitrary result if no codeword is found within distance $\delta\ell$. INSERT($x, S$) inserts a copy of $x$ into the multiset $S$, and DELETE($x, S$) deletes all copies of $x$ from $S$. FINDSIGNIFICANTENTRIES can be used in conjunction with DECOMPRESS to obtain a sparse approximation.

## 4.2 Details

We now fill in the details of the approach sketched in section 4.1. Consider the matrix sketches of (3). We use $p^{(t,r)}$ to denote polynomial $t$ in the sketch number $r$, for $r = 1, \ldots, 2\ell$.

For concreteness we consider an expander code [24], which is able to efficiently correct a fraction $\delta = \Omega(1)$ errors. Given a string within Hamming distance $\delta\ell$ from $E(x)$, the input $x$ can be recovered in time $\mathcal{O}(\ell)$, if the decoding algorithm is given access to a (fixed) lookup table of size $\mathcal{O}(n)$. (We assume without loss of generality that $\delta\ell$ is integer.)

Pseudocode for the algorithm computing the set of positions $(i,j)$ can be found in figure 3. For each splitting function $h^{(t)}(i,j) = (h_1^{(t)}(i) + h_2^{(t)}(j)) \bmod b$, and each hash value $k$ we try to recover any entry $(i,j) \in L \cap S_\Delta$ with $h^{(t)}(i,j) = k$. The recovery will succeed with good probability if there is a unique such entry. As argued in section 3.3 we get uniqueness for all but a small fraction of the splitting functions with high probability.

The algorithm first reads the relevant magnitude $X_r$ from each of the $2\ell$ sketches. It then produces a binary string $s$ that encodes which sketches have low and high magnitude (below and above $\Delta/2$, respectively). This string is then decoded into a pair of coordinates $(i, j)$, that are inserted in a multiset $S$.

A post-processing step removes "spurious" entries from $S$ that were not inserted for at least $d/2$ splitting functions, before the set is returned.

### 4.2.1 Proof of theorem 7

It is easy to see that FINDSIGNIFICANTENTRIES can be implemented to run in expected time $\mathcal{O}(db\ell)$, which is $\mathcal{O}(b \lg^2 n)$, and space $\mathcal{O}(db)$, which is $\mathcal{O}(b \lg n)$. The implementation uses the linear time algorithm for DECODE [24], and a hash table to maintain the multiset $S$.

Also, since we insert $db$ positions into the multiset $S$, and output only those that have cardinality $d/2$ or more, the set returned clearly has at most $2b$ distinct positions. It remains to see that each entry $(i, j) \in L \cap S_\Delta$ is returned with high probability. Notice that

$$((I_{E_r}A)B)_{ij} = \begin{cases} (AB)_{ij} & \text{if } E(i)_r = 1 \\ 0 & \text{otherwise} \end{cases}$$

$$(A(BI_{E_r}))_{ij} = \begin{cases} (AB)_{ij} & \text{if } E(j)_r = 1 \\ 0 & \text{otherwise} \end{cases}.$$

Therefore, conditioned on $h^{(t)}(i,j) = k$ we have that the random variable $X_r = |p^{(t,r)}[k]|$ has $\mathbf{E}[X_r] \in \{0, |(AB)_{ij}|\}$, where the value is determined by the $r$th bit of the string $\hat{s} = E(i)||E(j)$. The algorithm correctly decodes the $r$th bit if $X_r \leq \Delta/2$ for $\hat{s}_r = 0$, and $X_r > \Delta/2$ for $\hat{s}_r = 1$. In particular, the decoding of a bit is correct if $(AB)_{ij} \geq \Delta$ and $X_r$ deviates by at most $\Delta/2$ from its expectation.

From the proof of Theorem 6 we see that the probability that the error of a single estimator is greater than $12 \sqrt{\text{Err}_F^{b/20}(AB)/b}$ is at most $1/8$. If $\Delta$ is at least twice as large, this error bound implies correct decoding of the bit derived from the estimator, assuming $(AB)_{i^*j^*} > \Delta$. Adjusting constants 12 and 20 to a larger value $\kappa$ the error probability can be decreased to $\delta/3$. This means that the probability that there are $\delta\ell$ errors or more is at most $1/3$. So with probability $2/3$ DECODE correctly identifies $(i^*, j^*)$, and inserts it into $S$.

Repeating this for $d$ different hash functions the expected number of copies of $(i^*, j^*)$ in $S$ is at least $2d/3$, and by Chernoff bounds the probability that there are less than $d/2$ copies is $2^{-\Omega(d)}$. For sufficiently large $d = \mathcal{O}(\lg n)$ the probability that *any* entry of magnitude $\Delta$ or more is missed is less than $1/n$.

## 5. ESTIMATING COMPRESSIBILITY

To apply theorems 5 and 6 it is useful to be able to compute bounds on compressibility of $AB$. In the following subsections we consider, respectively, estimation of the number of nonzero entries, and of the $\text{Err}_F$ value.

## 5.1 Number of nonzero entries

An constant-factor estimate of $\bar{b} \geq b$ can be computed in time $\mathcal{O}(N \lg N)$ using Cohen's method [9] or its refinement for matrix products [3]. Recall that $\bar{b}$ is an upper bound on the number of nonzeros, when not taking into account that there may be zeros in $AB$ that are due to cancellation of terms. We next show how to take cancellation of terms into consideration, to get a truly output-sensitive algorithm.

The idea is to perform a doubling search that terminates (whp.) when we arrive at an upper bound on the number of nonzero entries that is within a factor $\mathcal{O}(1)$ from the true value. Initially, we multiply $A$ and $B$ by random diagonal matrices (on left and right side, respectively). This will not change the number of nonzero entries, but by the Schwartz-

Zippel lemma it ensures that a linear combination of entries in $AB$ is zero (whp.) only when all these entries are zero.

The doubling search creates sketches for $AB$ using $b = 2, 4, 8, 16, \ldots$ until, say, $\frac{4}{5}b$ entries of the sketch vector become zero for all hash functions $h^{(1)}, \ldots, h^{(d)}$. Since there is no cancellation (whp.), this means that the number of distinct hash values (under $h(i,j)$) of nonzero entries $(i,j)$ is at most $b/5$.

We wish to bound the probability that this happens when the true number $\tilde{b}$ of nonzero entries is larger than $b$. The expected number of hash collisions is $\binom{\tilde{b}}{2}/b$. If the number of distinct hash values of nonzero entries is at most $b/5$ the average number of collisions per entry is $\tilde{b}/(b/5) - 1$. This means that, assuming $\tilde{b} \geq b$, the observed number of collisions can be lower bounded as:

$$\tilde{b}\left(\frac{\tilde{b}}{b/5} - 1\right)/2 \geq \frac{\tilde{b}^2}{b/5}\frac{4}{5}/2 \geq \frac{2\tilde{b}}{\tilde{b}+1}\binom{\tilde{b}}{2}/b \ .$$

Note that the observed value is a factor $\frac{2\tilde{b}}{\tilde{b}+1} \geq 4/3$ larger than the expectation. So Markov's inequality implies that this happens with probability at most $3/4$. If it happens for all $d$ hash functions, we can conclude with probability $1 - (3/4)^{-d}$ that $b$ is an upper bound on the number of nonzero entries.

Conversely, as soon as $b/5$ exceeds the number $\tilde{b}$ of nonzero entries we are guaranteed to finish the doubling search. This means we get a 5-approximation of the number of non-zeros.

## 5.2 Upper bounding $\text{Err}_F^{b/\kappa}(AB)$

To be able to conclude that the result of FINDSIGNIFICANT-ENTRIES is correct with high probability, using theorem 7, we need an upper bound on $\sqrt{\text{Err}_F^{b/\kappa}(AB)/b}$. We will in fact show how to estimate the larger value

$$\sqrt{\text{Err}_F^0(AB)/b} = ||AB||_F/\sqrt{b},$$

so the allowed value for $\Delta$ found may not be tight. We leave it as an open problem to efficiently compute a tight upper bound on $\sqrt{\text{Err}_F^{b/\kappa}(AB)/b}$.

The idea is to make use of the AMS sketch $X$ of $AB$ using the approach described in section 2.1 (summing the sketches for the outer products). If we use 4-wise independent hash functions $s_1$ and $s_2$, Indyk and McGregor [18] (see also Braverman et al. [5] for a slight correction of this result) have shown that $X^2$ is an unbiased estimator for the second moment of the input, which in our case equals $||AB||_F^2$, with variance at most $8\mathbf{E}[X^2]^2$. By Chebychev's inequality this means that $X^2$ is a 32-approximation of $||AB||_F^2$ with probability $3/4$. (Arbitrarily better approximation can be achieved, if needed, by taking the mean of several estimators.) To make sure that we get an upper bound with high probability we take the median of $d = \mathcal{O}(\log n)$ estimators, and multiply this value by 32. The total time for computing the upper bound is $\mathcal{O}(dn^2)$.

## 6. CONCLUSION

We have seen that matrix multiplication allows surprisingly simple and efficient approximation algorithms in cases where the result is sparse or, more generally, its Frobenius norm is dominated by a sparse subset of the entries. Of course, this can be combined with known reductions of matrix multiplication to (smaller) matrix products [14, 23] to yield further (multiplicative error) approximation results.

**Acknowledgement.** The author thanks the anonymous reviewers for insightful comments, Andrea Campagna, Konstantin Kutzkov, and Andrzej Lingas for suggestions improving the presentation, Petros Drineas for information on related work, and Seth Pettie for pointing out the work of Iwen and Spencer.